\documentclass[%
 reprint,
 superscriptaddress,
showpacs,
 amsmath,amssymb,
 aps,
]{revtex4-1}

\usepackage{graphicx}
\usepackage{dcolumn}
\usepackage{bm}
\usepackage{amsmath}
\usepackage{hyperref}

\newcommand{\ket}[1]{|{#1}\rangle}
\newcommand{\bra}[1]{\langle{#1}|}
\newcommand{\ketbra}[2]{|{#1}\rangle\!\langle{#2}|}

\newcommand{\proj}[1]{\ketbra{#1}{#1}}
\newcommand{\Tr}{\text{Tr}}

\newcommand{\cL}{{\mathcal L}}
\newcommand{\cD}{{\mathcal D}}

\renewcommand{\Re}{\text{Re}}
\renewcommand{\Im}{\text{Im}}

\usepackage{color}

\begin{document}

\preprint{APS/123-QED}

\title{Quantum optical master equation for solid-state quantum emitters}

\author{Ralf Betzholz}
\affiliation{Theoretische Physik, Universit\"{a}t des
Saarlandes, D-66123 Saarbr\"{u}cken, Germany}
\author{Juan Mauricio Torres}
\affiliation{
Institut f\"ur Angewandte Physik, Technische Universit\"at Darmstadt, 
D-64289 Darmstadt, Germany
}

\author{Marc Bienert}
\affiliation{Theoretische Physik, Universit\"{a}t des
Saarlandes, D-66123 Saarbr\"{u}cken, Germany}

\date{\today}

\begin{abstract}
We provide an elementary description of the dynamics of defect centers in crystals in terms of a quantum optical master equation which includes spontaneous decay and a simplified vibronic interaction with lattice phonons. We present the general solution of the dynamical equation by means of the eigensystem of the Liouville operator and exemplify the usage of this damping basis to calculate the dynamics of the electronic and vibrational degrees of freedom and to provide an analysis of the spectra of scattered light. The dynamics and spectral features are discussed with respect to the applicability for color centers, especially for negatively charged nitrogen-vacancy centers in diamond.
\end{abstract}

\maketitle


\section{Introduction}

Nowadays quantum optics avails itself not only of purely atomic sources of quantum light. The demand for easy-to-handle sources of non-classical light for modern experiments and quantum technological applications triggered a wide search for alternatives in several fields of physics.
Among solid state systems, defect centers in diamond, primarily the negatively charged nitrogen-vacancy (NV$^-$) centers, have experienced an impressively successful development~\cite{nv:doherty2013} during the last decade. These defect centers provide optically addressable electronic transitions between discrete quantum states energetically situated between the valence and conduction band.  Anti-bunching of the emitted light as indicator for a single photon source was shown~\cite{kurtsiefer:2000}, Wheeler's delayed-choice experiment was realized~\cite{nv:jacques2007} and coupling to optical micro-cavities was performed~\cite{nv:albrecht2013} to mention a few of the experimental breakthroughs. Moreover, applicability of the centers for quantum information technology using coupled electronic and nuclear spins has been shown~\cite{dutt:2007,togan2010,waldherr2014}, and the defect centers can serve as high precision magnetic sensors~\cite{degen2008,balasubramanian2008,maze2008}.

In contrast to atoms, solid state quantum emitter do not need to be laser cooled and trapped, as they are fixed in their solid state matrix. However, this embedding usually leads to strong coupling of the electronic degrees of freedom to lattice vibrations. While single atoms ideally only couple to the free radiation field, resulting in a damping due to spontaneous decay, their solid state counterparts experience among other effects, additional dephasing or non-radiative decay due to the coupling with crystal phonons. In the atomic case, the damping is typically described by a Markovian master equation. Solid state systems lack such a handy theoretical description based on a simplified physical picture. With the contribution we develop here, we hope to participate in filling this gap.

One of the first, most elementary and quantitative model for color centers was developed by Huang and Rhys~\cite{nv:huang1950} based on the Franck-Condon principle applied to color centers in crystals. It models the vibrational modes all at the same frequency in a continuum approximation, hence being essentially a single mode concept. Later developments generalized the approach to an arbitrary vibronic spectrum~\cite{nv:lax1952,nv:maradudin1966, *nv:maradudin1967} and afterwards these ideas were applied to diamond~\cite{nv:davies1974}, especially for the calculation of absorption and emission spectra. More recent models include pragmatic approaches~\cite{nv:su2008}, using somehow artificially but successfully a certain number of vibrational levels associated with the electronic ground state, with each transition to the excited state damped by an individual Markovian bath. A description based on a non-Markovian master equation is reported in~\cite{nv:wilson-rae2002} for general phononic spectral functions.

The numerous applications, especially of NV$^-$ centers, in quantum photonics call for a tractable theoretical description of the dipole-phonon interaction which renders the basics of the photon's spectral properties and is based on a clear physical picture. In this work we will present such a model by combining the Franck-Condon principle with a master equation formalism.

The basic idea is to describe the coupling to the vibrational modes by a single harmonic oscillator. 
Many defects in lattices are accompanied by the presence of strongly coupled localized modes, which have a exponentially decaying~\cite{nv:maradudin1966, *nv:maradudin1967} amplitude away from the defect. These modes are represented by the single oscillator in our model. The delocalized phononic modes, in contrast, play the role of a temperature bath for the oscillator. We bring this idea into the form of a master equation and provide its full solution in terms of the damping basis, {\it i.e.} a spectral decomposition of the Liouville operator which generates the dynamics. In terms of this biorthogonal basis, we show how the absorption spectrum of a single defect center can be easily represented and analyzed. In such a way the basic features in the spectrum of light of solid state quantum emitters, in particular NV$^-$ centers, can be reproduced by a simple dynamical equation for the system's density operator. 

This paper has the following structure: In Sec.~\ref{sec:model} we present the model and set up the master equation. Sec.~\ref{sec:dampingbasis} is devoted to the damping basis as a solution of the dynamical equation. We derive the left- and right eigenelements of the Liouville operator together with its eigenvalues. In Sec.~\ref{sec:dynspec} we use the damping basis to describe the time evolution and calculate the absorption spectrum of a single defect center. We discuss the features of the model in Sec.~\ref{sec:discussion}, and put it into context with other descriptions. Finally, in Sec.~\ref{sec:conclusions} we draw the conclusions. In the appendix we provide some details and techniques used for the calculations in the main text.

\section{The model}
\label{sec:model}

In this section we develop a simplified theoretical description of a single color center coupled to a vibrational mode. This description employs the fact, that among the vibrational normal modes, often only a few localized modes strongly couple to the defect center, while delocalized phononic modes are considered here to play the role of a thermal bath causing the singled out localized mode to thermally relax.

\begin{figure}[tbp]
\begin{center}
\textbf{a)}\vtop{\vskip-2ex\hbox{\includegraphics[width=2.7cm]{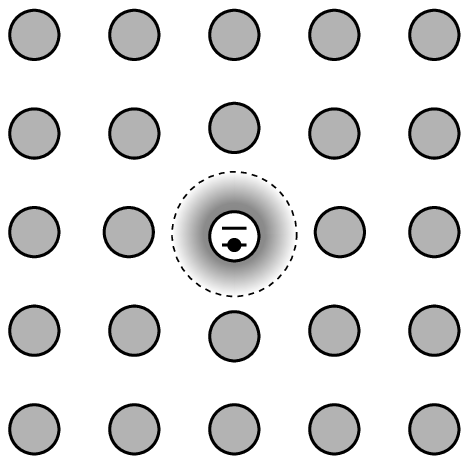}}}\hspace{2mm}
\textbf{b)}\vtop{\vskip-2ex\hbox{\includegraphics[width=2.7cm]{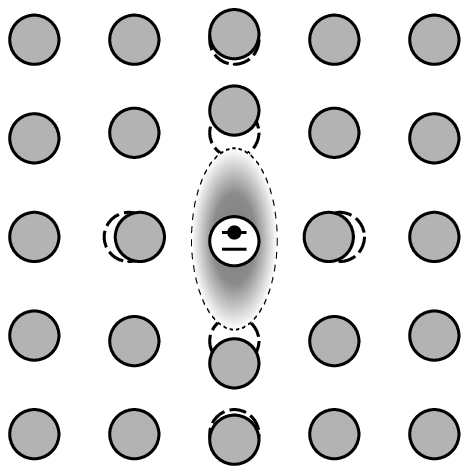}}}\\[6mm]
\textbf{c)}\vtop{\vskip-2ex\hbox{\includegraphics[width=4cm]{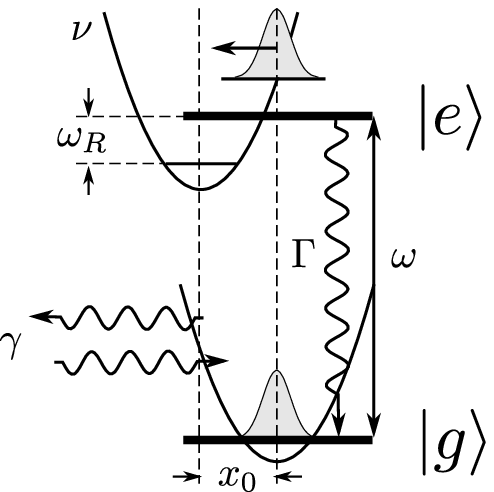}}}
\end{center}
\caption{\label{fig:model} a) The Coulomb interaction of the ground state electronic orbital (shaded gray) of a defect center lets the surrounding atoms take on a certain lattice position. b) The excited electronic orbital pushes the adjacent crystal atoms into a displaced equilibrium configuration. c) Model of the Franck-Condon like lattice-defect center interaction in single mode and harmonic approximation including damping: In the excited state $\ket{e}$, the wavepacket of the single vibrational mode experiences a potential displaced by $x_0$ compared to the potential in the ground state $\ket{g}$. The interaction causes a renormalized transition frequency between the states $\ket{e}$ and $\ket{g}$, shifted by the relaxation frequency $\omega_R$. The single vibrational mode is coupled to a temperature bath modeling the phononic modes of the crystal and thermally relaxes with rate $\gamma$. The electronic two-level system decays with rate $\Gamma$ due to spontaneous emission.}
\end{figure}

\subsection{Vibronic coupling}

We consider two electronic states with transition frequency $\omega$, a stable state $\ket{g}$ and an energetically higher state $\ket{e}$. For an NV$^-$ center these states can be identified with the $^3\text{A}_2$ and $^3\text{E}$ levels~\cite{nv:doherty2011}, respectively. The common understanding of the defect-vibration interaction assumes electronic-state dependent equilibrium positions of the atoms around the defect~\cite{nv:maradudin1966,*nv:maradudin1967}, see Fig.~\ref{fig:model}: Within the Franck-Condon assumption~\cite{franck1926, *condon1926}, atoms on the lattice sites around the defect experience a modified potential due to the reconfigured electronic orbitals, leading to an interaction term
\begin{align}
 V(x) =\sigma_g V_g(x) + \sigma_e V_e(x)
\end{align}
in the total Hamiltonian $H$ when only a single vibrational mode with coordinate $x$ is taken into account. Here the abbreviations $\sigma_g=\proj{g}$ and $\sigma_e=\proj{e}$ for the atomic projectors have been introduced for later convenience. For diamond, local vibration mode calculations~\cite{nv:gali2011} let it appear reasonable that only a few local modes strongly couple to the center, justifying the single mode assumption employed here. In the harmonic approximation and when expanding the potential 
\begin{align}
 V_e(x)&=V_g(x) + F x + \hbar\omega\nonumber\\
 &=V_g(x+x_0) + \hbar(\omega -\omega_R)\label{eq:Vclearform}
\end{align}
around its new equilibrium position $x_0 = F/M\nu^2$ up to first order in $x$, one finds
\begin{align}
 H &= \sigma_g H_\text{osc} +\sigma_e [\hbar \omega + H_\text{osc} + F x]\nonumber\\
 &= H_\text{osc} + H_\text{tls} +  W.
 \label{eq:H}
\end{align}
We defined the Hamiltonian 
\begin{align}
 H_\text{osc} = \frac{p^2}{2M}+\frac12 M \nu^2 x^2 = \hbar\nu \left[b^\dagger b + \frac12\right]
\end{align}
of a harmonic oscillator
with frequency $\nu$, momentum $p$ and effective mass $M$ associated with the local vibrational mode. The annihilation and creation operators $b$ and $b^\dagger$ are connected to the normal coordinate and momentum operators by
\begin{align}
 x &= \xi(b+b^\dagger),\\
 p &= \frac{\hbar}{2i\xi}(b-b^\dagger)
\end{align}
with the ground state length scale $\xi = \sqrt{\hbar/2M\nu}$. In the second line of Eq.~\eqref{eq:H}, we furthermore introduced the parts
\begin{align}
 H_\text{tls}& = \hbar\omega\sigma_e,\\
 W&=\hbar\eta\sigma_e(b+b^\dagger)
\end{align}
of the Hamiltonian 
representing the energy of the uncoupled two-level system and the interaction between the electronic and the vibrational degree of freedom with a coupling constant $\hbar\eta = F\xi$. Note that the coupling leads to an energy relaxation of $\hbar\omega_R=F^2/2M\nu^2$, what can already be seen in the form~\eqref{eq:Vclearform} of the interaction. Up to this point, the model corresponds basically to the description of Huang and Rhys~\cite{nv:huang1950,nv:doherty2013} of defect centers, describing the phononic degree of freedom by a single mode whose coordinate corresponds to the displacement of the surrounding lattice atoms from their equilibrium position. In this one-parametric model the dimensionless Huang-Rhys factor $S=\omega_R/\nu = (\eta/\nu)^2$ measures the vibronic interaction strength.

\subsection{Spontaneous decay and phononic damping}

To complete the description, we take electronic and vibrational relaxation processes into account using a master equation formalism. The time evolution of the system's density operator $\varrho$, covering the electronic and vibrational degree of freedom, reads
\begin{align}
 \frac{\partial\varrho}{\partial t} &= \cL\varrho\nonumber\\
 &=\frac{1}{i\hbar}[H, \varrho] + \cL_\Gamma \varrho + \cL_\gamma\varrho.
 \label{eq:me}
\end{align}
The first term on the right-hand side corresponds to the coherent evolution governed by the Hamiltonian $H$, Eq.~\eqref{eq:H}. Spontaneous decay with the rate $\Gamma$ is taken into account by the superoperator
\begin{align}
 \cL_\Gamma\varrho = \frac{\Gamma}{2}\,\cD[\sigma_-]\varrho,
\end{align}
where we defined the lowering operator $\sigma_-=\ketbra{g}{e}$ with the corresponding raising operator $\sigma_+=\ketbra{e}{g}$ and used the abbreviation  $\cD[X]\varrho = 2X\varrho X^\dagger-X^\dagger X\varrho-\varrho X^\dagger X$ to represent a Lindblad-form term. We further assume a finite lifetime of the vibrational excitation.  Imperfections in the lattice structure or non-harmonic corrections may lead to a coupling between the vibrational modes and let the mode $b$ interact with a bath of delocalized phononic modes with temperature $T$. In a Markovian approximation, such a coupling is described by
\begin{align}
 \cL_\gamma\varrho = \frac{\gamma}{2}(\bar m+1)\,\cD[b]\varrho + \frac{\gamma}{2}\bar m\,\cD[b^\dagger]\varrho
\end{align}
and damps the localized mode irreversibly towards its stationary state with mean vibrational occupation number $\bar m=[\exp(\hbar\nu/k_\text{B}T)-1]^{-1}$ on a time scale given by the inverse decay rate $1/\gamma$. Such a picture should be suitable for materials with few localized modes coupled strongly to the defect center compared to the delocalized modes forming the bath. The model also allows for a temperature dependent decay rate $\gamma(T)$, typical for many solid-state systems~\cite{nv:maradudin1962,freidkin1995,pouthier2010}. We furthermore remark that in the dissipation we neglected the dressing of the electronic states with the strongly interacting phonons of the oscillator, that leads to additional terms~\cite{hu2014} in the Liouvillian, Eq.~\eqref{eq:me}. Their influence can approximately be taken into account by an additional effective dephasing of the electronic subsystem.

\section{Solution of the master equation: Damping base}
\label{sec:dampingbasis}

An elegant form of providing a solution to the differential equation~\eqref{eq:me} is given in terms of the damping basis~\cite{qo:briegel1993}, namely the spectral decomposition of the Liouville operator $\cL$. To this end, we seek for the left and right eigenelements of the Liouvillian, fulfilling
\begin{align}
 \cL \hat\varrho_\lambda &= \lambda \hat\varrho_\lambda\label{eq:righteveq},\\
 \check\varrho_\lambda^\dagger \cL&= \lambda \check\varrho_\lambda^\dagger. 
\end{align}
The eigenelements are orthogonal by construction with respect to the scalar product $\Tr[\check\varrho_\lambda^\dagger\hat\varrho_{\lambda'}]=\delta_{\lambda,\lambda'}$. If they form a complete set, the closure relation can be written as
\begin{align}
 \sum_\lambda \hat\varrho_\lambda\otimes\check\varrho_\lambda = 1,
\end{align}
where the outer product is defined by its action $(\hat\varrho_\lambda\otimes\check\varrho_\lambda)X=\Tr[\check\varrho_\lambda^\dagger X]\hat\varrho_\lambda$ on an arbitrary operator $X$. The time evolution of an initial state $\varrho(t=0)$ can then be written as
\begin{align}
 \varrho(t) = \sum_\lambda c_\lambda e^{\lambda t} \hat\varrho_\lambda,
 \label{eq:rhot}
\end{align}
where the coefficients are given by $c_\lambda = \Tr[\check\varrho_\lambda^\dagger \varrho(t=0)]$. The remainder of this section is dedicated to the derivation of the explicit form of the damping basis of the problem at hand. The reader who is not interested in this technical part may skip the rest of this section without loss of coherence. The explicit formulas are summarized at the end of this section.

For later convenience we reformulate the master equation, Eq.~\eqref{eq:me}, by introducing the Liouvillians 
\begin{align}
 \cL_\text{tls}\varrho &= \frac1{i\hbar}\left[H_\text{tls},\varrho\right] + \cL_\Gamma\varrho,\\
 \cL_\text{osc}\varrho &= \frac1{i\hbar}\left[H_\text{osc},\varrho\right] + \cL_\gamma\varrho,
\end{align}
describing a two-level system underlying spontaneous decay and a damped harmonic oscillator, respectively. Additionally we introduce 
\begin{align}
 \cL_\text{int}\varrho &= \frac1{i\hbar}\left[W,\varrho\right] = \frac{\eta}{i}\left[\sigma_e (b+b^\dagger),\varrho\right]
\end{align}
and write the complete Liouvillian as 
\begin{align}
  \cL = \cL_\text{tls} + \cL_\text{osc} +\cL_\text{int}.
  \label{eq:L}
\end{align}
The superoperators $\cL_\text{tls}$ and $\cL_\text{osc}$ exclusively act on the electronic or vibrational subspace, respectively. Later on we will hark back to the damping basis of $\cL_\text{osc}$, which was derived in Ref.~\cite{qo:briegel1993}. For the sake of completeness, we list the explicit expressions here: The normalized eigenelements read
\begin{align}
 \hat\mu_{nl}(b,b^\dagger) =& \frac{(-1)^n}{(\bar m+1)^{|l|+1}}b^{\dagger(|l|+l)/2)}\nonumber\\
 &:\text{L}_n^{|l|}\left[\frac{b^\dagger b}{\bar m\!+\!1}\right]\exp\left[-\frac{b^\dagger b}{\bar m\!+\!1}\right]:b^{(|l|-l)/2},\label{eq:muhatosc}\\
 \check\mu^\dagger_{nl}(b,b^\dagger) =& \left[\frac{-\bar m}{\bar m +1}\right]^n\frac{n!}{(n+|l|)!}b^{\dagger(|l|-l)/2)}\nonumber\\
 &:\text{L}_n^{|l|}\left[\frac{b^\dagger b}{\bar m}\right]:b^{(|l|+l)/2}\label{eq:mucheckosc}
\end{align}
(where colons $:...:$ denote normal ordering, $\text{L}_n^{l}(x)$ the associated Laguerre polynomials) with their corresponding eigenvalues
\begin{align}
 \lambda_{nl} = -i l\nu -\left[n+\frac{|l|}2\right]\gamma, \quad &n=0,1,2,\dots,\; \label{eq:lambdaosc}\\&l=0,\pm1, \pm2, \dots\nonumber
\end{align}
These eigenelements fulfill the completeness relation
\begin{align}
 \sum_{n=0}^\infty\sum_{l=-\infty}^\infty \hat\mu_{nl}\otimes\check\mu_{nl} = 1.\label{eq:mucr}
\end{align}
In the following, we will express the eigenbasis of the coupled system in terms of these eigenelements of a damped quantum harmonic oscillator.

\subsection{Right eigenelements}
\label{sec:relements}

Inspired by the damping basis of a two-level system with spontaneous decay~\cite{jakob2003}, the form of the Liouvillian, Eq.~\eqref{eq:me}, suggests the ansatz
\begin{align}
\hat\varrho_{\lambda}^{(0)} &= \sigma_g\,\hat\mu_{\lambda}^{(0)},\label{eq:rho0ansatz}\\
\hat\varrho_{\lambda}^{(\pm)} &= \sigma_\pm\,\hat\mu_{\lambda}^{(\pm)},\label{eq:rhopmansatz}\\
\hat\varrho_{\lambda}^{(\rightsquigarrow)}&= \sigma_e \hat\mu_{\lambda}^{(e)} - \sigma_g \hat\mu_{\lambda}^{(g)}\label{eq:rhodansatz}
\end{align}
for the right eigenelements. Here, $\mu$ are operators of the oscillator only and the eigenvalues $\lambda$ depend on $n$, $l$ and the superscript index of Eqs.~\eqref{eq:rho0ansatz}-\eqref{eq:rhodansatz}.

\subsubsection{Eigenelements connected to the atomic populations}

We begin with Eq.~\eqref{eq:rhodansatz}, since it will also deliver the result for Eq.~\eqref{eq:rho0ansatz}. Plugging the ansatz into Eq.~\eqref{eq:righteveq} and using $\cL$ in the form~\eqref{eq:L} yields
\begin{align}
  [\cL-\lambda]
  \hat\varrho_{\lambda}^{(\rightsquigarrow)} =& 
  \sigma_e\left\{[\cL_{\rm osc}^{(\beta)}-\Gamma
  -\lambda]\hat\mu_{\lambda}^{(e)}\right\}\nonumber\\
  -&\sigma_g\left\{[\cL_\text{osc}-\lambda]
  \hat\mu_{\lambda}^{(g)}-\Gamma\hat\mu_{\lambda}^{(e)}\right\}=0,
 \label{eq:dummy0}
\end{align}
where we defined the Liouvillian of a displaced harmonic oscillator
\begin{align}
 \cL_\text{osc}^{(\beta)}\mu =& \cL_\text{osc}\mu -i\eta\left[(b+b^\dagger),\mu\right]\nonumber\\
 =&D(\beta)\left[\cL_\text{osc}D^\dagger(\beta)\mu D(\beta)\right]D^\dagger(\beta)
 \label{eq:dispLosc}
\end{align}
with the displacement operator $D(\beta)=\exp[\beta b^\dagger-\beta^\ast b]$ taking 
\begin{align}
\beta = \frac{-\eta}{\nu-i\gamma/2}. \label{eq:beta}
\end{align}
This quantity will appear throughout the rest of the treatment of this problem and is connected to a generalization of the Huang-Rhys parameter.

One possibility to fulfill Eq.~\eqref{eq:dummy0} is to choose 
$\hat\mu_\lambda^{(e)}=0$.
Then only a component in the $\sigma_g$-subspace remains, obeying 
\begin{align}
  \cL_\text{osc}\hat\mu_{\lambda}^{(0)} = \lambda\hat\mu_{\lambda}^{(0)}.
  \label{eq:eveq0}
\end{align}
This obviously is the case from Eq.~\eqref{eq:rho0ansatz}, and we consequently relabeled the operator in Eq.~\eqref{eq:eveq0} correspondingly. The solution is given by Eq.~\eqref{eq:muhatosc} and the eigenvalues are the ones from Eq.~\eqref{eq:lambdaosc}. 

We now turn to the case $\hat\mu_{\lambda}^{(e)}\neq 0$. From Eqs.~\eqref{eq:dummy0} and~\eqref{eq:dispLosc} it follows immediately
\begin{align}
  \hat\mu_{\lambda}^{(e)} &= D(\beta)\hat\mu_{nl}D^\dagger(\beta),\\
  \hat\mu_{\lambda}^{(g)} &=  
  \frac{\Gamma}{\cL_\text{osc}-\lambda_{nl}+\Gamma}\hat\mu_\lambda^{(e)}
\end{align}
with the eigenvalues 
$
\lambda = \lambda_{nl}-\Gamma.
$

\subsubsection{Eigenelements connected to the atomic coherences}

It remains to deal with the eigenelements associated with the electronic coherences, Eqs.~\eqref{eq:rhopmansatz}. The action of $\cL$ yields
\begin{align}
  [\cL\!-\!\lambda]\sigma_\pm\hat\mu_{\lambda}^{(\pm)} 
  &\!=\!\sigma_\pm
  \left\{\cL_\text{osc}\!+\!\cL_\pm\mp i\omega\!-\!\frac\Gamma 2\!-\!\lambda\right\}
  \hat\mu_{\lambda}^{(\pm)}
  =0.
 \label{eq:dummy1}
\end{align}
Here, we introduced
$\cL_\pm\mu = i\eta([\mu,b+b^\dagger]\mp\{\mu,b+b^\dagger\})/2$.
To handle the eigenvalue equation, it is helpful to transform the expressions according to
\begin{align}
  \hat\mu_{\lambda}^{(\pm)} = 
  D(\alpha_\pm) \hat{\tilde\mu}_{\lambda}^{(\pm)}D^\dagger(\beta_\pm)
\label{eq:transalphabeta}
\end{align}
and choose the displacement parameters $\alpha_\pm$ and $\beta_\pm$ such that $\cL_\pm\mu\to [b,\mu]$, at the cost of an additional constant term. This turns out to be possible if one sets
\begin{alignat}{2}
 \alpha_+ &= \beta(\bar m+1)-\beta^\ast\bar m,\label{eq:alphap}\\
 \beta_+ &= (\beta-\beta^\ast)(\bar m+1),\\
 \alpha_- &= (\beta^\ast-\beta)\,\bar m,\\
 \beta_- &= \beta^\ast(\bar m+1) -\beta\bar m.
 \label{eq:betam}
\end{alignat}
The condition on the curly bracket of Eq.~\eqref{eq:dummy1} then transforms into the eigenvalue equation
\begin{align}
  \cL_\text{osc}&\hat{\tilde\mu}_{\lambda}^{(\pm)} \mp (2\bar m+1)\beta\gamma \left[b,\hat{\tilde\mu}_{\lambda}^{(\pm)}\right]
  =\left(\lambda\pm i\widetilde\omega+\frac{\widetilde\Gamma}2\right)\hat{\tilde\mu}_{\lambda}^{(\pm)}
 \label{eq:dummy3}
\end{align}
with the renormalized atomic transition frequency and linewidth
\begin{align} 
 \widetilde\omega &= \omega -|\beta|^2\nu,\label{eq:renomega}\\
 \widetilde\Gamma &=\Gamma + |\beta|^2\gamma(2\bar m+1).\label{eq:renGamma} 
\end{align}
In App.~\ref{app:Qfunc} it is shown that the lopsided $\hat\mu_{nl}(b,b^\dagger\pm\varsigma)$, with $b^\dagger$ shifted by  
\begin{align}
 \varsigma = (\beta^\ast-\beta)(2\bar m+1),
 \label{eq:varsigma}
\end{align}
are the right eigenelements of the operator on the left-hand side of Eq.~\eqref{eq:dummy3} to the eigenvalue  $\lambda_{nl}$, Eq.~\eqref{eq:lambdaosc}. The 
$\hat\mu_{\lambda}^{(\pm)}$ are found with the help of the transformation~\eqref{eq:transalphabeta} and the eigenvalues
\begin{align}
  \lambda = \lambda_{nl}\mp i\widetilde\omega-\frac{\widetilde\Gamma}2.
\end{align}
are composed of the harmonic oscillator part and the renormalized electronic transition frequency and linewidth.

\subsection{Left eigenelements}
The derivation of the left eigenelements of $\cL$ goes along the same lines as for their right counterparts. The action to the left~\cite{qob:englert:five_lectures} of $\cL$ is represented by
\begin{align}
 \check\varrho_\lambda^\dagger \cL = &\frac{i}{\hbar}\left[H,\check\varrho_\lambda^\dagger\right]
 +\frac{\Gamma}2\check\cD[\sigma_-]\check\varrho_\lambda^\dagger\nonumber\\&+\frac{\gamma}2(\bar m+1)\check\cD[b]\check\varrho_\lambda^\dagger+\frac{\gamma}2\bar m\check\cD[b^\dagger]\check\varrho_\lambda^\dagger,
\end{align}
where the short notation $\check\cD[b]X = 2b^\dagger X b- b^\dagger b X -X b^\dagger b$ was introduced. This time
\begin{align}
  \check\varrho_{\lambda}^{(0)\dagger} &= 
  \sigma_g\,\check\mu_{\lambda}^{(g)\dagger}+\sigma_e\,\check\mu_{\lambda}^{(e)\dagger},\label{eq:rhoc0ansatz}\\
  \check\varrho_{\lambda}^{(\pm)\dagger} &= \sigma_\mp\,\check\mu_{\lambda}^{(\pm)\dagger},\label{eq:rhocpmansatz}\\
  \check\varrho_{\lambda}^{(\rightsquigarrow)\dagger} &= \sigma_e \check\mu_{\lambda}^{(\rightsquigarrow)\dagger} \label{eq:rhocdansatz}
\end{align}
serves as an ansatz. Following an analogous treatment as in Sec.~\ref{sec:relements}, we start with Eq.~\eqref{eq:rhoc0ansatz}, as it will again deliver the solution for Eq.~\eqref{eq:rhocdansatz} too. One arrives at
\begin{align}
  \check\mu_{\lambda}^{(g)\dagger} &=\check\mu_{nl}^\dagger\\
  \check\mu_{\lambda}^{(e)\dagger} &=D(\beta)\left[D^\dagger(\beta)\check\mu_{nl}^\dagger D(\beta)\frac{\Gamma}{\Gamma\!+\!\lambda_{nl}\!-\!\cL_\text{osc}}\right]D^\dagger(\beta)
\end{align}
to the eigenvalue $\lambda=\lambda_{nl}$ and
\begin{align}
 \check\mu_{\lambda}^{(\rightsquigarrow)\dagger} &=D(\beta)\check\mu_{nl}^\dagger D^\dagger(\beta)
\end{align}
to the eigenvalue $\lambda = \lambda_{nl}-\Gamma$. 
Also the electronic coherence parts can be derived similarly as before, but this time using the transformation
\begin{align}
  \check\mu_{\lambda}^{(\pm)\dagger} = D(\beta_\pm)\check{\tilde\mu}_{\lambda}^{(\pm)\dagger}D^\dagger(\alpha_\pm)
\end{align}
with $\alpha_\pm$ and $\beta_\pm$ from Eqs.~\eqref{eq:alphap}-\eqref{eq:betam}. One finally obtains
\begin{align}
  \check\mu_{\lambda}^{(\pm)\dagger} = D(\beta_\pm)\check\mu_{nl}^\dagger(b,b^\dagger\pm\varsigma)D(\alpha_\pm)
\end{align}
for the left eigenelements.

\subsection{Summary}%
In the previous subsections we found that the eigenelements of the model Liouvillian $\cL$, Eq.~\eqref{eq:L}, are assigned to the stable electronic subspace $\sigma_g$, the electronic coherences $\sigma_\pm$ and a subspace which can be associated with the population decay in the excited state at rate $\Gamma$,
\begin{align}
  \hat\varrho_{nl}^{(0)} =& \sigma_g\,\hat\mu_{nl}(b,b^\dagger), \label{eq:rho0sum}\\
  \hat\varrho_{nl}^{(\pm)} =& \sigma_\pm\,D(\alpha_\pm)\hat\mu_{nl}(b,b^\dagger\pm\varsigma)D^\dagger(\beta_\pm), \\
  \hat\varrho_{nl}^{(\rightsquigarrow)} =& \left\{\sigma_e\!
  -\!\sigma_g\,\Gamma[\cL_\text{osc}\!-\!\lambda_{nl}\!+\!\Gamma]^{-1} \right\}D(\beta)\hat\mu_{nl}(b,b^\dagger)D^\dagger(\beta).\label{eq:hatrhodecay}
\end{align}
They fulfill the duality relation 
$\Tr[\check\varrho_{nl}^{(\jmath)\dagger}\hat\varrho_{n'l'}^{(\jmath')}]=
\delta_{\jmath,\jmath'}\delta_{n,n'}\delta_{l,l'}$ ($\jmath,\jmath'\in \{0,\pm,\rightsquigarrow\}$)
together with their left-side counterparts 
\begin{align}
\check\varrho_{nl}^{(0)\dagger}  =& \sigma_g\, \check\mu^\dagger_{nl}(b,b^\dagger) +\sigma_e\, \Gamma D(\beta)\left\{D^\dagger(\beta)\check\mu_{nl}^\dagger(b,b^\dagger) D(\beta)\right. \nonumber\\
&\left.\quad\quad\quad\quad\quad\times[\Gamma\!+\!\lambda_{nl}\!-\!\cL_\text{osc}]^{-1}\right\}D^\dagger(\beta) \label{eq:check0sum},\\
\check\varrho_{nl}^{(\pm)\dagger}  =& \sigma_\mp\, D(\beta_\pm)\check\mu_{nl}^\dagger(b,b^\dagger\pm\varsigma)D^\dagger(\alpha_\pm),\\
\check\varrho_{nl}^{(\rightsquigarrow)\dagger}  =&\sigma_e\, D(\beta)\check\mu_{nl}^\dagger(b,b^\dagger)D^\dagger(\beta),
\end{align}
both having the common eigenvalues
\begin{align}
  \lambda_{nl}^{(0)} &= \lambda_{nl}\label{eq:lambda0sum},\\
  \lambda_{nl}^{(\pm)} &=\lambda_{nl}\mp i\widetilde\omega-\frac{\widetilde\Gamma}{2}\label{eq:lambdapmsum},\\
  \lambda_{nl}^{(\rightsquigarrow)} &= \lambda_{nl}-\Gamma\label{eq:lambdasquigsum}
\end{align}
for non-negative integers $n$ and arbitrary integers $l$. The renormalized frequency $\widetilde\omega$ and linewidth $\widetilde\Gamma$ of the two-level system are
given by Eqs.~\eqref{eq:renomega} and~\eqref{eq:renGamma}, respectively. Moreover, $\alpha_\pm$, $\beta_\pm$ and $\varsigma$ are defined in Eqs.~\eqref{eq:alphap}-\eqref{eq:betam} and Eq.~\eqref{eq:varsigma} in terms of the model parameter $\beta$, Eq.~\eqref{eq:beta}. The stationary state $\varrho_{\rm st}=\hat\varrho_{00}^{(0)} = \sigma_g \mu_{\rm th}$ of the system corresponds to $\lambda=0$ and describes the atom in the ground state and the vibrational mode in thermal equilibrium, {\it i.e.} $\mu_{\rm th}=\hat\mu_{00}(b,b^\dagger)$. The corresponding left eigenelement $\check\varrho_{00}^{(0)}=1$ is the identity operator. We remark that the formal representations~\eqref{eq:hatrhodecay} and~\eqref{eq:check0sum} containing $\cL_\text{osc}$ can be made explicit by inserting the completeness relation of the $\hat\mu_{nl}$ together with the relation~\eqref{eq:Crel}. Additional pure electronic dephasing of the form $\Gamma^\ast\mathcal{D}[\sigma_e]/2$ can easily be incorporated with the replacement $\widetilde\Gamma\rightarrow \widetilde\Gamma+\Gamma^\ast$ in Eq.~\eqref{eq:lambdapmsum}, while all the eigenelements remain unchanged.

\section{Application of the damping basis: Dynamics and spectra}
\label{sec:dynspec}

\subsection{Time evolution}
\label{sec:dynamics}

The dynamics of an arbitrary initial state in terms of the damping basis is given by Eq.~\eqref{eq:rhot}. We focus on the case where the oscillator is initially in thermal equilibrium and the two-level system is prepared in an arbitrary state $\rho_0$: $\varrho(t=0) = \rho_0 \mu_\text{th}$. For the time evolution of the reduced density operator $\rho(t)=\Tr_\text{osc}[\varrho(t)]$ describing the dynamics of the atom, one finds
\begin{align}
\rho(t) = 
\sum_{n=0}^\infty\sum_{l=-\infty}^\infty \Tr_\text{osc}\Big[&
 c_{nl}^{(0)} \hat\varrho_{nl}^{(0)}e^{\lambda_{nl} t} + c_{nl}^{(\rightsquigarrow)} \hat\varrho_{nl}^{(\rightsquigarrow)}e^{(-\Gamma+\lambda_{nl}) t}\nonumber\\
 +&\sum_{\jmath=\pm}c_{nl}^{(\jmath)}\hat\varrho_{nl}^{(\jmath)}e^{\lambda_{nl}^{(\jmath)} t}\Big]
\label{eq:redrhot}.
\end{align}
The expansion coefficients 
$c_{nl}^{(\jmath)}=\Tr[\check\varrho_{nl}^{(\jmath)\dagger}\varrho(t=0)]$ 
are determined by the initial state $\varrho(t=0)$ and are calculated using the right eigenelements 
$\check\varrho_{nl}^{(\jmath)\dagger}$ with $\jmath\in\{0,\pm,\rightsquigarrow\}$.
The partial trace in Eq.~\eqref{eq:redrhot} over the oscillator degrees of freedom yields
\begin{align}
 \Tr_\text{osc}[\hat\varrho_{nl}^{(0)}] &= \sigma_g\delta_{n,0}\delta_{l,0},\label{eq:nulldeltanl}\\
 \Tr_\text{osc}[\hat\varrho_{nl}^{(\rightsquigarrow)}] &= (\sigma_e-\sigma_g)\delta_{n,0}\delta_{l,0},\label{eq:squigdeltanl}
\end{align}
according to Eqs.~\eqref{eq:rho0sum} and~\eqref{eq:hatrhodecay}. For the first one this becomes clear when recalling that the only right oscillator eigenelement with non-vanishing trace is the stationary state $\hat\mu_{00}(b,b^\dagger)$. The same argument can be used to show the second one, after inserting the completeness relation~\eqref{eq:mucr} next to the superoperator in the eigenelement~\eqref{eq:hatrhodecay}. For the two coherences, one finds
\begin{align}
  \Tr_\text{osc}[ \hat\varrho_{nl}^{(\pm)}] &= \sigma_\pm A_{nl}^\pm, \label{eq:trpm}
\end{align}
where we defined the trace over the oscillator operators in $\hat\varrho_{nl}^{(\pm)}$ as
\begin{align}
A_{nl}^\pm = \Tr[D(\alpha_\pm)\hat\mu_{nl}(b,b^\dagger\pm\varsigma)D^\dagger(\beta_\pm)].\label{eq:Anlpmdef}
\end{align}
The evaluation of these traces is conveniently carried out using phase space distributions, as it is sketched in App.~\ref{app:A}. The resulting functions depend only on $\beta$ and $\bar m$ and are presented in formulas~\eqref{eq:Aplus}-\eqref{eq:Aminus}.

We now turn to the expansion coefficients in Eq.~\eqref{eq:redrhot}. Because of Eqs.~\eqref{eq:nulldeltanl}-\eqref{eq:squigdeltanl}, for the two coefficients in the first line we only need the $n=l=0$ case,
\begin{align}
 c_{00}^{(0)} &= \Tr[(\sigma_g+\sigma_e)\rho_0]= 1,\\
 c_{00}^{(\rightsquigarrow)} &= \Tr[\sigma_e\rho_0]= \rho_{ee},
 \end{align}
giving unity and the initial excited state population $\rho_{ee}=\bra e\rho_0\ket{e}$. In the second line of Eq.~\eqref{eq:redrhot}, the coefficients for the coherences
\begin{align}
c_{nl}^{(\pm)} &= \Tr[\sigma_\mp\rho_0] B_{nl}^\pm = \rho_\pm B_{nl}^\pm\label{eq:cpm}
\end{align}
are required, where we have defined the initial values $\rho_\pm = \Tr[\sigma_\mp\rho_0]$ and
\begin{align}
B_{nl}^\pm=\Tr[D(\beta_\pm)\check\mu_{nl}^\dagger(b,b^\dagger\pm\varsigma)D(\alpha_\pm)^\dagger\mu_{\rm th}].
\end{align}
Again we use phase space representations of the required eigenelements to evaluate the trace. The details are reported in App.~\ref{app:B}. It is clear from the form of the solution~\eqref{eq:redrhot} that
\begin{align}
A_{nl}^- B_{nl}^- = &\exp\left[{-\bar m\beta^{2}-(\bar m+1)\beta^{\ast 2}}\right]\frac{[\bar m(\bar m+1) |\beta|^4]^{n}}{n!(n+|l|)!} \nonumber\\&
\times\left\{
 \begin{array}{cl}
   \left[(\bar m+1)\beta^{\ast 2}\right]^{|l|}&\;\text{for}\quad l<0\\
   \left[ \bar m\beta^2\right]^{l}&\;\text{for}\quad l\ge 0
 \end{array}
\right.\label{eq:ABmin}
\end{align}
and  $A_{nl}^+ B_{nl}^+ =  [A_{n,-l}^- B_{n,-l}^-]^\ast$ only appear as products. As a matter of course, for $t=0$ the initial state must be reproduced by the solution~\eqref{eq:redrhot} of the master equation. Hence, if the eigensystem is complete, the remaining sum over $n$ and $l$ has to fulfill this requirement. For finite times, the exponentials containing the eigenvalues generate the dynamics. The form of the eigenvalues is linear in $n$ and $l$, such that the time dependency can be easily incorporated into Eq.~\eqref{eq:ABmin}. The sums can then be performed with the help of the power series representation
\begin{align}
 \left(\frac{x}{2}\right)^l\sum_{n=0}^\infty \frac{(x/2)^{2n}}{n!(n+l)!}=\text{I}_{l}(x)
\end{align}
and the generating function of the modified Bessel function $\text{I}_l(x)$ of order $l$~\cite{math:abramowitz1964}. After performing these steps one arrives at the solution
\begin{align}
 \rho(t) = &\sigma_g + \rho_{ee}(\sigma_e - \sigma_g) e^{-\Gamma t} \nonumber\\ +&\left[\rho_-\sigma_-(t)\exp\left\{\bar m \bar\beta^2(t)+(\bar m+1)\bar\beta^{2\ast}(t)\right\}+\text{H.c.}\right]\label{eq:rhot2}
\end{align}
with the time-dependent quantities
\begin{align}
 \bar\beta^2(t) &= \beta^2\left(e^{-[i\nu+\frac{\gamma}2] t}-1\right),\label{eq:betat}\\
 \sigma_\pm(t) &=\sigma_\pm e^{\mp i\widetilde\omega t-\widetilde\Gamma t}\label{eq:sigmat}.
\end{align}
The expression~\eqref{eq:rhot2} indeed reproduces the initial electronic state for $t=0$, as expected. For finite times, the population in the excited state decays with the natural rate $\Gamma$ of the two-level system. It is the coherence between the ground and excited state which is modified by the coupled phonon mode. 
This influence on the temporal behavior is twofold: First, the operators $\sigma_\pm$, Eq.~\eqref{eq:sigmat}, rotate and decay with the renormalized frequencies from Eq.~\eqref{eq:lambdapmsum}. Second, the evolution is modulated by the time-dependency of $\bar\beta^2(t)$, Eq.~\eqref{eq:betat}. The oscillator hence generates an effective dephasing of the two-level system.

In many-body systems the vibrational degree of freedom can also play a crucial role~\cite{pouthier2008,silva2006}, we therefore briefly delve into its dynamics in order to provide additional insight into the overall temporal behavior. We focus on the reduced density operator of the oscillator, $\mu(t) = \Tr_\text{tls}[\varrho(t)]$ with the same initial state $\varrho(t=0)$ as above. The dynamics is described by Eq.~\eqref{eq:redrhot} with the trace this time going over the electronic degrees of freedom, leaving only the terms in the first line. With the help of the completeness relation~\eqref{eq:mucr} and the overlaps of displaced harmonic oscillator eigenelements calculated in App.~\ref{app:C}, the expression for $\mu(t)$ can be recast into a form, where the
 time-dependent exponential factors can be incorporated into the displacement parameter
\begin{align}
 \beta(t) = \beta e^{-[i\nu +\frac{\gamma}2]t}.
\end{align}
Then, the time evolution of the oscillator's state can be expressed as
\begin{align}
 \mu(t) = \rho_{gg}\mu_\text{th} &+\rho_{ee} e^{-\Gamma t} D(\beta-\beta(t))\mu_\text{th} D^\dagger(\beta-\beta(t))\nonumber\\
 +\rho_{ee}\Gamma &\int\limits_0^t {\rm d}\tau \, e^{-\Gamma\tau} D(\beta(t-\tau)-\beta(t))\nonumber\\
 &\times \mu_\text{th} D^\dagger(\beta(t-\tau)-\beta(t)).\label{eq:mut}
\end{align}
The result consists of three contributions: (i) The thermal state corresponding to the $\ket{g}$-part of $\varrho(t=0)$ showing no dynamics. (ii) A dynamically displaced thermal state component moving in the $\ket{e}$ potential surface, whose population dies off with the electronic decay rate and (iii) the flow of the decayed population towards the undisplaced thermal state. In the asymptotic limit $t\to\infty$ the oscillators state $\mu(t)$ ends in the thermal state $\mu_\text{th}$, as expected.

\begin{figure}
\begin{center}
\textbf{a)}
\vtop{\vskip-1ex\hbox{\includegraphics[width=0.4\textwidth]{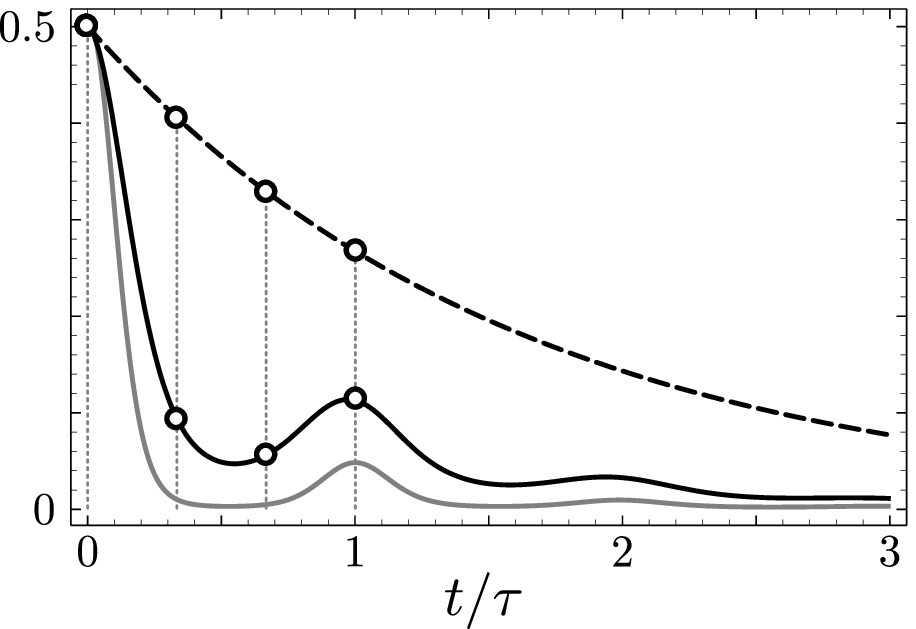}}}\\[4mm]
\textbf{b)}
\vtop{\vskip-1ex\hbox{\includegraphics[width=0.4\textwidth]{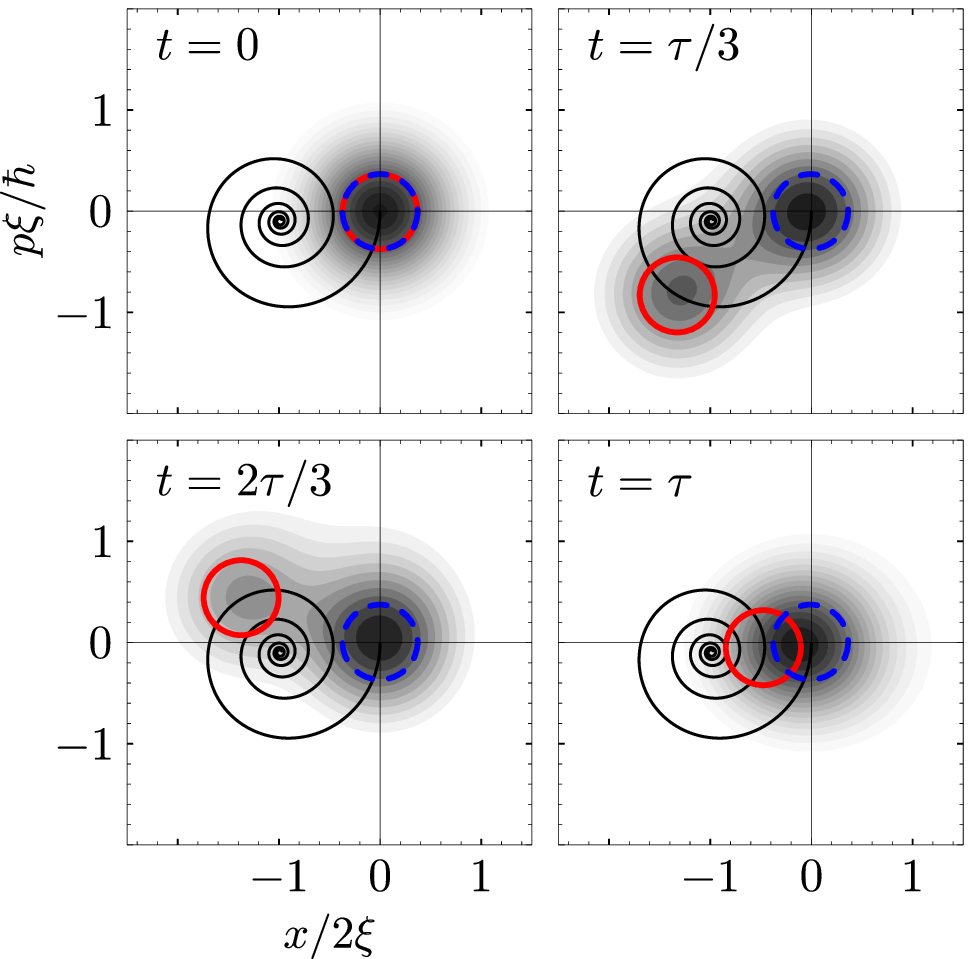}}}
\end{center}
\caption{\label{fig:dynamics}(Color online) Part a) shows the time evolution of the excited state population (dashed line) and the modulus of the coherences (solid lines) for the initial atomic state $(\vert g\rangle+\vert e\rangle)/\sqrt{2}$ in units of the oscillator period $\tau=2\pi/\nu$ when the oscillator is initially in thermal equilibrium. The parameters are $\gamma=0.2\nu$, $\Gamma=0.1\nu$ and $\eta=1\nu$. The black curves show the case $\bar m=0.05$ while the gray one corresponds to $\bar m=1$. The density plots b) show the Wigner function of the oscillator state at the times indicated by the circles in a). The dashed blue ring represents the variance and location of the $\proj{g}$-projection of the oscillator state in phase space, whereas the solid red ring analogously represents the $\proj{e}$-projection which follows a trajectory along the spiral around $\beta$. The excited state population dies off and the system ultimately ends up in the stationary state $\proj{g}\mu_\text{th}$.}
\end{figure}

In Fig.~\ref{fig:dynamics} we illustrate the dynamics of both the electronic and the vibrational degree of freedom.
The initial state of the two-level system is $\rho_0 = \proj{\psi_0}$ with $\ket{\psi_0} = (\ket{g}+\ket{e})/\sqrt{2}$ and  a thermal state for the oscillator. Subplot a) shows the time evolution of the population in the excited state (dashed curve) which decays exponentially with rate $\Gamma$, as in the unperturbed two-level system. The time evolution of the coherences (black solid curve) is modified by the oscillator: We show the absolute value of $\bra{e}\rho(t)\ket{g}$ which exhibits a faster decay compared to the pure two-level case, superimposed by an oscillating modulation with frequency $\nu$ as described by Eq.~\eqref{eq:rhot2}. For larger temperature (gray solid curve) the coherences are damped away even faster. Fig.~\ref{fig:dynamics}b) visualizes the temporal behavior of the oscillator's quantum state, Eq.~\eqref{eq:mut}, in phase space with the help of the Wigner function~\cite{qob:schleich:quantum_optics_phase_space}. We show four snapshots of the Wigner function within one oscillation period $\tau=2\pi/\nu$ at the times marked in subfigure a) by empty circles. Initially, at $t=0$, the phase space distribution corresponds to a thermal state, {\it i.e.} a Gaussian with variance proportional to $\bar m+1/2$. The portion of the distribution corresponding to the excited state starts to spirally circulate around the displaced equilibrium position at $x/2\xi=\Re\,\beta$ and $p\xi/\hbar=\Im\,\beta$ and its population decays simultaneously at rate $\Gamma$, as described in the second term of Eq.~\eqref{eq:mut}. The black spiral visualizes the trajectory of the center of probability of this part while the blue (dashed) and red (solid) rings mark the variance of the Gaussian components moving in the harmonic potential of the ground and excited state, respectively. The contribution of the last term of Eq.~\eqref{eq:mut} is hardly visible, but results, for example at $t=2\tau/3$, in a small asymmetry with respect to the axis connecting the centers of the two Gaussians. At $t=\tau$, the components moving in the ground and excited state potentials strongly overlap  again, leading to the partial revival of the electronic coherence in Fig.~\ref{fig:dynamics}a).

\subsection{Absorption and emission spectrum}
The advantage of the eigenvalue decomposition of the Liouvillian becomes apparent most clearly when calculating the spectral properties of the system. The method we use here was already applied for a systematic analysis of the fluorescence light of laser cooled atoms~\cite{bienert2004} and is now carried over to the case of solid state quantum emitters. For a weak probe laser of frequency $\omega_\text{L}$ illuminating the defect center, the spectrum of absorption under stationary conditions is determined by~\cite{qo:mollow1972,qo:cohen1977}
\begin{align}
  \mathcal{A}(\omega_\text{L}) = \Re\int\limits_0^\infty \, {\rm d}t\, \langle [\sigma_-,\sigma_+(t)]\rangle_\text{st}e^{-i\omega_\text{L} t}.
 \label{eq:S}
\end{align}
In this form the correlation function is evaluated without the probe. The time evolution of the operators is determined by the Liouvillian $\cL$, Eq.~\eqref{eq:L}, and can be calculated using the quantum regression theorem~\cite{qob:carmichael:statistical_methods1} to rewrite $\langle[\sigma_-,\sigma_+(t)]\rangle_\text{st} = \Tr[\sigma_+ e^{\mathcal L t}[\varrho_{\rm st},\sigma_-]]$. The stationary state of the undriven system is $\varrho_\text{st}=\proj{g}\mu_\text{th}$, and hence, the second term of the commutator vanishes such that the spectrum takes on the form 
\begin{align}
  \mathcal{A}(\omega_\text{L}) &= \Re\int\limits_0^\infty \, {\rm d}t\, \Tr[\sigma_+ e^{\cL t}\varrho_\text{st}\sigma_-]e^{-i\omega_\text{L} t}\label{eq:S1}\\
  &= \Re\sum_\lambda \frac{1}{i\omega_\text{L}-\lambda} \Tr[\sigma_+\hat\varrho_\lambda]\Tr[\check\varrho^\dagger_\lambda\mu_\text{st}\sigma_-]. \label{eq:S2}
\end{align}
In the last step we expanded the spectrum in terms of the eigensystem of $\cL$, Eqs.~\eqref{eq:rho0sum}-\eqref{eq:lambdasquigsum}, and performed the integration.  The traces in Eq.~\eqref{eq:S2} are readily calculated, we encountered them already in the previous subsection. They give
\begin{align}
 \Tr[\sigma_+\hat\varrho_{nl}^{(-)}]\Tr[\check\varrho^{(-)\dagger}_{nl}\mu_\text{st}\sigma_-] = A_{nl}^{-} B_{nl}^{-},
\end{align}
with the explicit form of $A_{nl}^{-}$ and $B_{nl}^{-}$ specified in App.~\ref{app:traces} and the product $W_{nl}=A_{nl}^{-}B_{nl}^{-}$ shown in Eq.~\eqref{eq:ABmin}. 
The decomposed spectrum
\begin{align}
 \mathcal{A}(\omega_\text{L}) &=\Re\sum_{n,l} \frac{W_{nl}}{i\omega_\text{L}-\lambda^{(-)}_{nl}}
 \label{eq:absspec}
\end{align}
is a superposition of curves associated with each eigenvalue $\lambda_{nl}^{(-)}$, weighted by the function $W_{nl}$ which also determines the shape: For real $W_{nl}$ the spectral contribution is a Lorentzian, for purely imaginary weight it is a Fano-like profile, and in the general case it is a mixture of both, positioned at $\omega_\text{L} = \Im\,\lambda_{nl}^{(-)}$ with a width given by $\Re\,\lambda_{nl}^{(-)}$. From the $l$ dependency it follows that the spectrum is built up by a group of peaks separated by the mechanical frequency $\nu$ and situated around the renormalized transition frequency $\widetilde\omega$. The peaks are distinguishable for $\Gamma,\gamma\ll\nu$ and otherwise overlap. Each peak is a sum of several contributions of increasing width for rising $n$, whereby the $n>0$ contributions vanish when $\bar m\to0$. In this limit the weight factors take on the form
\begin{align}
  W_{nl}=\delta_{n,0}\frac{(\beta^{\ast 2})^{|l|}}{|l|!} e^{-\beta^{\ast 2}}\label{eq:WnlT0}
\end{align}
 for $l\le 0$ and otherwise zero. It goes over into a Poissonian distribution when $\beta$ becomes real. For vanishing thermal occupation of the phonon mode, only phonons can be created during the absorption process and consequently, the spectral components for $l>0$ disappear. Note that the absorption and excitation spectrum of the two-level system coincide in the situation discussed here. We also remark that we had arrived to the same result by calculating the action of the propagator $e^{\cL t}$ in Eq.~\eqref{eq:S1} to the right using the solution~\eqref{eq:rhot2}. The integral representation can be advantageous in some cases, for example in the high temperature limit, when the expansion, Eq.~\eqref{eq:S2}, converges slowly.
 
\begin{figure}[tbp]
\begin{center}
\includegraphics[width=0.4\textwidth]{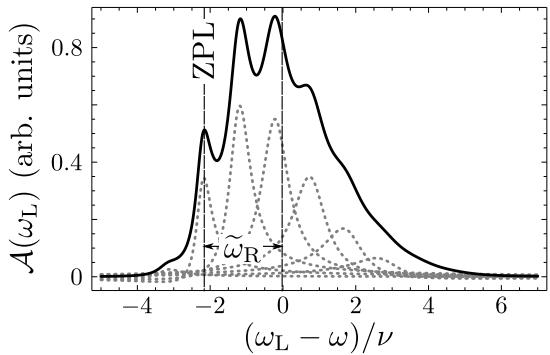}
\end{center}
\caption{\label{fig:absorption1} The absorption spectrum of the defect center under weak excitation, according to Eq.~\eqref{eq:absspec}, for the parameters $\gamma=0.2\nu$, $\Gamma=0.01\nu$, $\eta=1.5\nu$ and $\bar m=0.05$. The gray dotted curves show the main contributions of the eigenvalue decomposition~\eqref{eq:absspec} of the spectrum, namely the terms corresponding to $n=0$ and $l=-5,...,1$. ZPL denotes the position of the zero-phonon line shifted by $\widetilde\omega_R$ from the bare transition frequency of the electronic states.}
\end{figure}
 
In Fig.~\ref{fig:absorption1} we show the absorption spectrum, Eq.~\eqref{eq:absspec}, as a function of the probe frequency $\omega_\text{L}$ for a temperature close to zero.
The spectrum is composed by several resolved peaks. The zero-phonon line corresponding to a pure electronic transition without creation or annihilation of phononic excitations is shifted by the relaxation frequency $\widetilde\omega_R=\widetilde\omega-\omega$ from the frequency of the bare transition $\ket{e}\leftrightarrow\ket{g}$. 
More peaks are visible at multiple distances of the oscillator frequency and extend towards higher frequencies, forming the vibronic sidebands corresponding to absorption processes accompanied by the creation of phonons. 
Phonons to be annihilated are only rarely present at low temperatures entailing strongly suppressed signals on the lower frequency side of the zero-phonon line. 
The dashed curves show the most relevant terms of the sum in Eq.~\eqref{eq:absspec}, that is the individual spectral components associated with a certain eigenvalue which compose together the total spectrum. Note that the single contributions can become negative, only the superposition generally yields a valid spectrum.

We conclude this section by discussing the spectral properties of a photon emitted from the defect center. The emission spectrum reads~\cite{qob:scully:quantum_optics}
\begin{align}
 {\mathcal E}(\omega_\text{p}) =  \Re\int\limits_0^\infty {\rm d}t\int\limits_0^\infty {\rm d}\tau\, \langle\sigma_+(t+\tau)\sigma_-(t)\rangle e^{-i\omega_\text{p} \tau}
\end{align}
We assume here that the two-level system is initially excited and the oscillator is in the corresponding equilibrium $\mu_0 = D(\beta)\mu_\text{th}D^\dagger(\beta)$. This is a reasonable assumption if the lifetime of the excited state $\Gamma^{-1}$ is much larger than the thermalization time scale $\gamma^{-1}$ of the phonon mode. Again the quantum regression theorem and the completeness relation of the eigenelements are applied to bring the spectrum into the form
\begin{align}
 {\mathcal E}(\omega_\text{p}) = \Re\sum_{\lambda,\lambda'}\frac{\Tr[\sigma_+\hat\varrho_\lambda]\Tr[\check\varrho_\lambda^\dagger\sigma_-\hat\varrho_{\lambda'}]\Tr[\check\varrho_{\lambda'}^\dagger \sigma_e\mu_0]}{(\lambda-i\omega_\text{p})\lambda'}.\label{eq:esepcdecomp}
\end{align}
The further evaluation uses
\begin{align}
 \Tr[\sigma_+\hat\varrho_\lambda] &= A_{nl}^-,\\
 \Tr[\check\varrho_{\lambda'}^\dagger \sigma_e\mu_0] &= \delta_{n',0}\delta_{l',0}\Tr[\check\varrho^{(\rightsquigarrow)\dagger}_{n'l'}\sigma_e\mu_0],
\end{align}
and thus $\lambda'=-\Gamma$. The remaining trace in Eq.~\eqref{eq:esepcdecomp} can be calculated using the techniques of App.~\ref{app:traces} and yields $(-1)^l B_{nl}^+e^{-i \Im\beta^2}$, {\it viz.} Eq.~\eqref{eq:Bplus}.
The spectrum 
\begin{align}
 {\mathcal E}(\omega_\text{p}) = \frac{1}{\Gamma}\Re\sum_{n,l}\frac{W'_{nl}}{i\omega_\text{p}-\lambda^{(-)}_{nl}}
\end{align}
contains the weight factors $W'_{nl}=W_{n,-l}^\ast$, signifying the fact that the emission spectrum is a mirrored version of the absorption spectrum~\cite{mahan2000} with respect to the modified transition frequency $\widetilde\omega$.

\section{Discussion}
\label{sec:discussion}

We modeled the vibronic interaction of defect centers in crystals using a single-mode description including the radiative damping of the electronic degree of freedom as well as the damping of the vibrational mode by a heat bath with a temperature corresponding to a mean vibrational quantum number $\bar m$. The model can therefore be considered as an extension of the artificial but instructive single-mode Hamiltonian description, sometimes called the Huang-Rhys model~\cite{nv:doherty2013} or Einstein model~\cite{mahan2000}, which is the limiting case of our model when both linewidths, $\Gamma$ and $\gamma$, tend to zero.

From the temporal behavior treated in Sec.~\ref{sec:dynamics} follows that the vibronic coupling results in an effective dephasing mechanism for the electronic subsystem: It leaves the dynamics of the population unaffected but expedites the decay of the electronic coherences. This decay of the coherences is superimposed by an oscillatory behavior whose origins can be traced back to the motion of the vibrational degree of freedom in Wigner phase space. In the spirit of master equations, the dynamics of the two-level system can be considered as being dephased by a structured environment with harmonic time evolution, when tracing out the vibrational coordinates. The different types of dynamics concerning the populations and coherences is formally resembled in the classes of eigenvalues, Eqs~\eqref{eq:lambda0sum}-\eqref{eq:lambdasquigsum}: The excited state simply decays with the natural rate $\Gamma$. In contrast, the electronic coherences oscillate with a renormalized frequency $\widetilde\omega=\omega-\widetilde\omega_R$ and decay with the increased, temperature-dependent rate $\widetilde\Gamma=\Gamma+\widetilde\Gamma_R$. 
The relaxation frequency $\widetilde\omega_R=|\beta|^2\nu$ goes over into  $\omega_\text{R}=S\nu$, proportional to the Huang-Rhys factor $S$, when the linewidth $\gamma$ tends to zero. 
For finite linewidth, the damping rate $\gamma$ additionally modifies the transition frequency of the two-level system, reflecting the fact that the electronic subsystem indirectly couples to the Markovian environment which damps the oscillator. Compared to the uncoupled case, the decay rate is increased by 
$\widetilde\Gamma_R= |\beta|^2\gamma(2\bar m+1)$.

\begin{figure}[tbp]
\begin{center}
\textbf{a)}\vtop{\vskip-1ex\hbox{\includegraphics[width=0.4\textwidth]{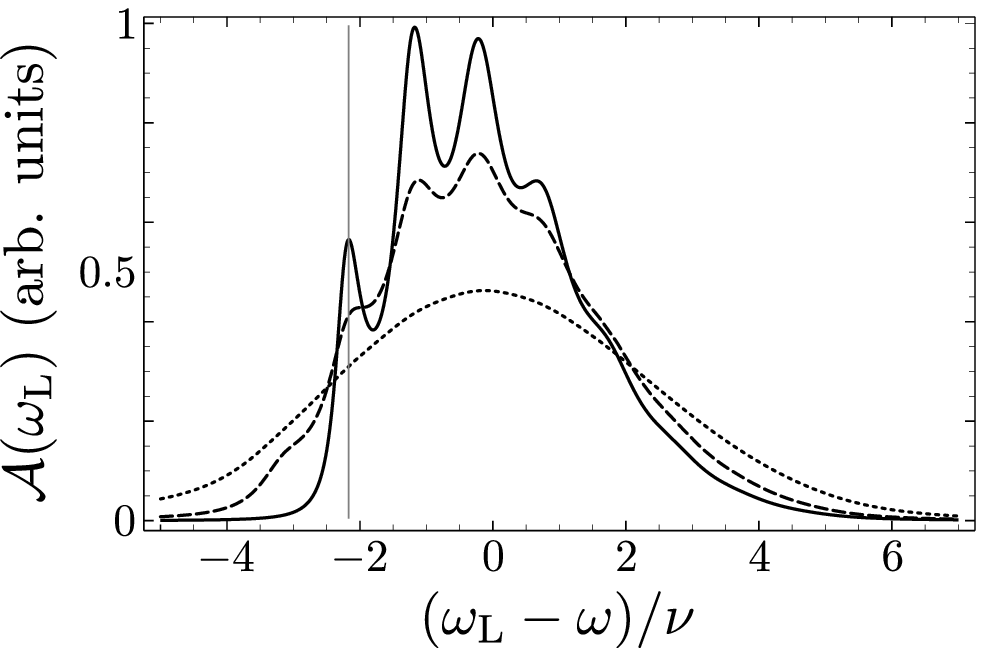}}}\\[5mm]
\textbf{b)}\vtop{\vskip-1ex\hbox{\includegraphics[width=0.4\textwidth]{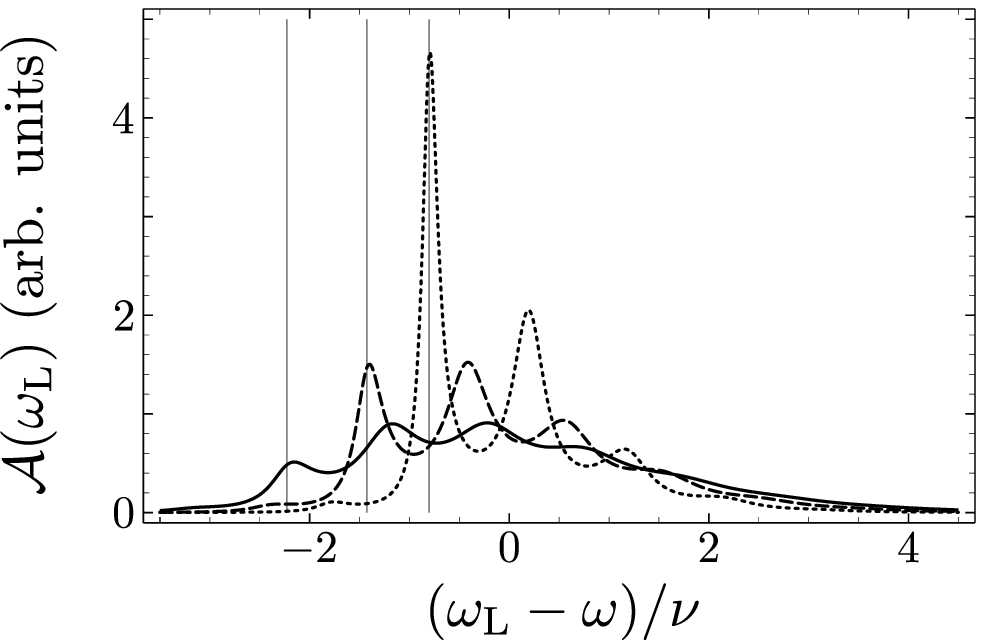}}}
\end{center}
\caption{\label{fig:abs} a) The temperature dependence of the absorption spectrum for $\bar m=0$ (solid line), $\bar m=0.25$ (dashed line) and $\bar m=1$ (dotted line). b) Dependence of the absorption on the coupling strength for $\eta=1.5\nu$ (solid line),  $\eta=1.2\nu$ (dashed line) and $\eta=0.9\nu$ (dotted line). The vertical lines are drawn at the corresponding zero-phonon lines, {\it i.e.} at the renormalized atomic frequency $\widetilde\omega$. The other parameters are the same as in Fig.~\ref{fig:absorption1}. 
}
\end{figure}

Concerning the spectra, again a comparison with the Huang-Rhys model suggests itself. According to Ref.~\cite{nv:davies1981}, this single-mode Hamiltonian model in the harmonic and Franck-Condon approximation, results in an absorption spectrum consisting of Dirac-$\delta$-shaped spikes displaced from the renormalized frequency $\omega-\omega_R$ by multiples of the vibrational frequency $\nu$, whose envelope (related to the intensities) is described by a Poissonian distribution $I_\ell=e^{-S} S^\ell/\ell!$, when $\ell$ numbers the vibronic sidebands. The generalized formulation applied here reflects these characteristics formally in Eq.~\eqref{eq:WnlT0}, with the Huang-Rhys parameter replaced by the complex parameter $\beta^{\ast2}$ quantifying the vibronic interaction strength. For finite temperature, the Poissonian-like distribution breaks down and the intensity of the $l$-th vibronic sideband relative to the total spectral intensity is described by
\begin{align}
 I_l = e^{-\Re\beta^2\coth x}\,\text{I}_l\left(\frac{|\beta|^2}{\sinh x}\right)e^{-l x}\cos\theta_l
\end{align}
with $\theta_l=\Im\,\beta^2+\arg\beta^{2l}$ and $x=\hbar\nu/2k_B T$. This is a meaningful expression when the vibronic sidebands are resolved, that is for $\gamma,\Gamma\ll\nu$. The dependency of the absorption spectrum on the temperature is depicted in Fig.~\ref{fig:abs}a). For low temperature $\bar m\approx 0$ (solid line), the spectrum consists of the zero phonon line at $\omega_\text{L}=\omega-\widetilde\omega_R$ with $\widetilde\omega_R\approx 2\nu$ for the parameters considered here. 
towards higher frequencies the vibronic sidebands are stretched out, starting to mutually overlap due to the swelling width of increasing orders. For higher temperature, $\bar m=0.25$, the individual lines are only hardly distinguishable and extend also towards lower frequencies on the red side of the zero-phonon line. The position of the zero phonon line is not temperature dependent, in contrast to its width given by $\Re\,\lambda^{(-)}_{n0}$, Eq.~\eqref{eq:lambdapmsum}. Finally for $\bar m = 1$ only a broad spectral shape is present, centered approximately at the bare transition frequency $\omega$ of the two-level system. For $\nu\gg\gamma,\Gamma$ this single peak approaches a Gaussian form around $\omega_\text{L}=\omega$ whose width is proportional to $\sqrt{\bar m}$. Fig.~\ref{fig:abs}b) shows the absorption spectrum at low temperature for three different vibronic coupling strengths, $\eta=0.9\nu$, $1.2\nu$ and $1.5\nu$, corresponding to the dotted, dashed and solid curves, respectively. The peak strength follows the quasi-Poissonian distribution~\eqref{eq:WnlT0}, with more peaks appearing for larger coupling: then higher order phonon processes become more likely. The position of the zero-phonon line moves towards lower frequencies for larger couplings in accordance with the renormalized frequency $\widetilde\omega$ defined in Eq.~\eqref{eq:lambdapmsum} and the simplified physical picture of Fig.~\ref{fig:model}. Moreover, the curves in Fig.~\ref{fig:abs}b) demonstrate that an increased coupling goes along with a broadening of the spectral components, as described by $\widetilde\Gamma$, Eq.~\eqref{eq:lambdapmsum}.

The master equation description of this work can also be related to the independent boson model~\cite{mahan2000}, in which a linear coupling to essentially a continuum of phononic modes is taken into account. The phase diffusion in the temporal evolution of the phononic modes is at the heart of the dephasing mechanism there, being replaced in the present description by a Markovian bath. Consequently, the spectral function describing the frequency dependent coupling strength of the phononic continuum in that model is replaced here by a single parameter, namely the damping rate $\gamma$ of the singled out phononic mode. Depending on the intended application, the more detailed independent boson model or the one-parametric simplified description presented here can be advantageous.

Finally we comment on the applicability of the model for NV$^-$ centers in diamond. As already mentioned, the presence of a few localized modes of similar frequencies suggests to highlight these modes -- or ultimately a single mode -- as it was done here. A temperature bath damping the mode has clear physical origins in the continuum of the delocalized phononic modes. The Markovian description, however, is a simplification which works reasonably well for rendering the basic properties of the NV$^-$ dynamics, but does not include the details of the temperature dependency of the zero-phonon line, for example. This would at least require to replace the flat spectral function implicitly applied here, if not more involved processes~\cite{nv:fu2009} have to be included, depending on the desired accuracy of the description. A temperature-dependent damping rate $\gamma(T)$ can however be used to mimic such behaviour. Of course, effects stemming from the dynamical Jahn-Teller effect are absent. Nevertheless, the model renders the basic properties of the spectra of NV$^-$ centers, especially in the room temperature regime and goes beyond simple pure dephasing models or artificial fitting models. Note that the room temperature spectra still correspond to $\bar m\ll 1$ due to the high Debye-temperature of diamond.

\section{Conclusions}
\label{sec:conclusions}

We formulated a compact theoretical description for the dynamics of defect centers in crystals. This master equation description is based on a linear vibronic coupling in the Franck-Condon, single phonon mode and two-level approximation and includes radiative and phononic Markovian damping. The overall solution of the master equation is given in terms of the eigenelements and eigenvalues of the Liouville operator. This solution covers both, the electronic and phononic degree of freedom and can be further reduced to a master equation of the electronic two-level system only by tracing out the undesired degree of freedom. We demonstrated the calculation of the absorption and emission spectrum at different temperatures, which take on a insightful form and reproduce the basic features of real spectra, for example for NV$^-$ centers in diamond. The physical picture at the heart of the presented model was reviewed in detail by means of the dynamics of both involved degrees of freedom.

In the form presented here, the model can be easily extended by adding more vibrational modes in order to increase the accuracy of the theory, if desired. Moreover, such an extension would also be interesting from the point of view of a systematic approximation of open quantum dynamics~\cite{woods2014}. It is also entirely conceivable to extend the treatment to include quadratic vibronic coupling or more electronic levels.

We believe that the model, based on an easy but clear physical picture, will be helpful for investigations of defect centers in quantum optical applications, for example as single photon sources, when the influence of the vibronic sidebands becomes important. With the analytic solution provided here it can deliver a basis for the description of coupled cavity-defect-center systems or similar hybrid setups.  

\section{Acknowledgments}
We thank Giovanna Morigi, Christoph Becher and Ignacio Wilson-Rae for helpful comments and discussions. Furthermore, M.B. and R.B. acknowledge support from the German Research Foundation (DFG) within the project BI1694/1-1 and from the GradUS program of the Saarland University. J.M.T. acknowledges support by CONACyT
(Mexico) Postdoctoral Grant No. 162781, by the BMBF
project Q.com, and by CASED III.

\begin{appendix}
 \section{Calculation of eigenelements}
 \label{app:Qfunc}
In Sec.~\ref{sec:relements} we encountered the eigenvalue equation of the shifted superoperator
\begin{align}
 \cL_\text{shift}\tilde\mu=\cL_\text{osc}\tilde\mu\pm c\left[b,\tilde\mu\right] = \lambda\tilde\mu
 \label{eq:eveqLshift}
\end{align}
for which we now derive the right eigenelements and eigenvalues $\lambda$. For this purpose we use the $Q$-representation~\cite{qob:schleich:quantum_optics_phase_space} of the eigenelements,
\begin{align}
 Q(\beta,\beta^\ast) = \frac{1}\pi\bra{\beta}\tilde\mu\ket{\beta},
\end{align}
defined as the expectation value of $\tilde\mu$ in a coherent state $\ket{\beta}$. In terms of this phase space function, Eq.~\eqref{eq:eveqLshift} is represented by the partial differential equation
\begin{align}
 \left[L_\text{FP} \pm c\frac{\partial}{\partial\beta^\ast}\right]Q(\beta,\beta^\ast) = \lambda Q(\beta,\beta^\ast).
\end{align}
with the differential operator
\begin{align}
 L_\text{FP}(\beta,\beta^\ast)=&\left(i\nu+\frac\gamma2\right)\frac{\partial}{\partial \beta}\beta+\left(-i\nu+\frac\gamma2\right)\frac{\partial}{\partial \beta^\ast}\beta^\ast\nonumber\\+&\gamma(\bar m+1)\frac{\partial^2}{\partial\beta\partial\beta^\ast}
\end{align}
connected to the Fokker-Planck equation of an Ornstein-Uhlenbeck process~\cite{qob:risken:fokkerplanck}, describing the motion of a damped harmonic oscillator in $Q$-phase space~\cite{qob:carmichael:statistical_methods1}. With the help of the coordinate transformation $\bar\beta^\ast = \beta^\ast\pm\varsigma$ with $\varsigma=c/(-i\nu+\gamma/2)$, the extra term can be removed, and one finds the standard form
\begin{align}
 L_\text{FP}(\beta,\bar\beta^\ast)\bar Q(\beta,\bar\beta^\ast) = \lambda \bar Q(\beta,\bar\beta^\ast). 
\end{align}
The solution leads -- after the substitution $\beta\to b$, $\bar\beta^\ast\to b^\dagger$ in normal order -- to the eigenelements  $\hat\mu_{nl}$, Eq.~\eqref{eq:muhatosc}~\cite{qo:briegel1993}. In our case, first the coordinate transformation is reversed, and eventually
\begin{align}
 \hat\mu_{nl}(b,b^\dagger\pm\varsigma)
\end{align}
results as the right eigenelement of Eq.~\eqref{eq:eveqLshift}, with eigenvalues given by $\lambda_{nl}$, Eq.~\eqref{eq:lambdaosc}.
Similarly the corresponding left eigenelements can be derived using the action to the left of $\cL_{\rm shift}$.

\section{Overlaps between eigenelements}
\label{app:traces}
In Sec.~\ref{sec:dynspec} we need explicit expressions for the formulas
\begin{align}
A_{nl}^\pm &= \Tr[D(\alpha_\pm)\hat\mu_{nl}(b,b^\dagger\pm\varsigma)D^\dagger(\beta_\pm)]~\label{eq:appAnl},\\
B_{nl}^\pm &=\Tr[D(\beta_\pm)\check\mu_{nl}^\dagger(b,b^\dagger\pm\varsigma)D^\dagger(\alpha_\pm)\mu_{\rm th}]\label{eq:appBnl}
\end{align}
following from overlaps of eigenelements of the Liouvillian $\cL$, Eq.~\eqref{eq:L}. Moreover,
\begin{align}
C_{nl}^{mk} = \Tr&[\check\mu_{nl}^\dagger D(\beta)\hat\mu_{mk}D^\dagger(\beta)]\label{eq:appCnlmk}
\end{align}
is helpful for the application of the damping basis.

\subsection{The traces $A_{nl}^\pm$}
 \label{app:A}
To evaluate expression~\eqref{eq:appAnl}, we use the cyclic property of the trace and
\begin{align}
 e^{\mp\varsigma b}D^\dagger(\beta_\pm)D(\alpha_\pm)e^{\pm \varsigma b}=D(\pm\beta^\ast)\exp[\mp\beta\varsigma]e^{i\Im \alpha_{\pm}\beta^\ast_{\pm}}.
\end{align}
One arrives at
\begin{align}
 A_{nl}^\pm = e^{i\Im \alpha_{\pm}\beta^\ast_{\pm}}e^{-\frac12|\beta|^2}\chi_{nl}(\mp\beta^\ast)
\end{align}
with the characteristic function
\begin{align}
 \chi_{nl}(\beta) &= \Tr\left[D(\beta)e^{\frac12|\beta|^2}\hat\mu_{nl}(b,b^\dagger)\right]\nonumber\\
 &=\int\mathrm{d}^2\alpha\, e^{\alpha\beta^\ast-\alpha^\ast\beta} {Q}_{nl}(\alpha)
\end{align}
of antinormally ordered expectation values~\cite{glauber1969}, being the Fourier transform of the quasi-probability distribution $Q_{nl}(\alpha)=\bra{\alpha}\hat\mu_{nl}(b,b^\dagger)\ket{\alpha}$ which is easily obtained from Eq.~\eqref{eq:muhatosc}. The remaining integral can be evaluated using the generating function
\begin{align}
 \frac{\exp[\frac{zx}{z+1}]}{(z+1)^{l+1}}=\sum_{n=0}^\infty \text{L}_n^l(x) (-z)^n\label{eq:lgf}
\end{align}
of the Laguerre polynomials. After collecting all terms, one finds
\begin{align}
A_{nl}^{+}=&\frac{(-1)^n}{n!} [(\bar m+1) |\beta|^2]^{n}\left\{
 \begin{array}{c}
  \beta^{\ast |l|}\\
 (-\beta)^{l}
 \end{array}
\right\}\nonumber\label{eq:Aplus}\\
&e^{-(\bar m+1) \text{Re}\beta^{2}}e^{-\bar m\beta^{\ast 2}}e^{\left(\bar m+\frac{1}{2}\right) |\beta|^2},\\
A_{nl}^{-}=&\frac{(-1)^n}{n!} [(\bar m+1) |\beta|^2]^{n}\left\{
 \begin{array}{c}
  (-\beta^\ast)^{|l|}\\
  \beta^{l}
 \end{array}
\right\}\nonumber\label{eq:Aminus}\\
&e^{-\bar m(\text{Re}\beta^{2}+\beta^{\ast 2})}e^{-\beta^{\ast 2}}e^{\left(\bar m+\frac{1}{2}\right) |\beta|^2}.
\end{align}
In this notation, the upper and lower line in the curly brackets correspond to $l<0$ and $l\ge 0$, respectively.

\subsection{The traces $B_{nl}^\pm$}
\label{app:B}
Equation~\eqref{eq:appBnl} has the phase space representation
\begin{align}
 B_{nl}^\pm = \pi\int {\rm d}^2\alpha\, \check P_{nl}(\alpha) Q(\alpha)
\end{align}
with 
\begin{align}
 \check P_{nl}(\alpha) &= \frac{(-1)^n}{\pi}\frac{n!}{(n+|l|)!}\left\{\begin{array}{c}\!\alpha^{\ast|l|}\\\alpha^{l}\!\end{array}\right\}\text{L}_n^{|l|}\left(\frac{|\alpha|^2}{\bar m\!+\!1}\right),\\
 Q(\alpha) &= \frac1\pi\bra{\alpha}e^{\pm\varsigma b}D(\alpha_\pm)\mu_\text{th}D(\beta_\pm)e^{\mp\varsigma b}\ket{\alpha}.\label{eq:QB}
\end{align}
The first line is the $P$-distribution of the left eigenelements Eq.~\eqref{eq:mucheckosc}. All the other operators we collected in the $Q$-function, Eq.~\eqref{eq:QB}, including the thermal state $\mu_\text{th}$, where $\ket{\alpha}$ denotes a coherent state. The $Q$-function can be evaluated explicitly by applying the Baker-Hausdorff relation and using the properties of coherent states. Then, in the function
\begin{align}
 \widetilde{B}_l^{\pm}(z) = \sum_{n=0}^\infty \frac{(n+|l|)!}{n!} B_{nl}^\pm z^n
 \label{eq:wB}
\end{align}
the summation is performed using Eq.~\eqref{eq:lgf}, leading to an integral of the form
\begin{align}
 \int {\rm d}^2\alpha \left\{\begin{array}{c}\alpha^{\ast|l|}\\\alpha^{l}\end{array}\right\} e^{\alpha\zeta^\ast-\alpha^\ast\xi}e^{-c|\alpha|^2} = \frac{\pi}{c^{l+1}}\left\{\begin{array}{c}\zeta^{\ast|l|}\\(-\xi)^l\end{array}\right\}e^{-\frac{\xi\zeta^\ast}{c}}.\label{eq:intform}
\end{align}
For the ``$-$''-case, the coefficients are 
\begin{align}
 \xi &= -\frac{\bar m}{\bar m+1}\beta,\quad \zeta^\ast = -\beta^\ast,\\
 c^{-1}&=(z+1)(\bar m+1).
\end{align}
Finally, one obtains
{
\allowdisplaybreaks
\begin{align}
B_{nl}^{+}=&\frac{(-1)^n}{(n+|l|)!}[\bar m |\beta|^{2}]^n 
\left\{
 \begin{array}{c}
  [\bar m\beta^\ast]^{|l|}\\
  \left[-(\bar m+1) \beta\right]^l
 \end{array}
\right\}\nonumber\\*
&e^{-i(\bar m+1) \Im \beta^2}e^{-\left(\bar m+\frac{1}{2}\right) |\beta|^2},\label{eq:Bplus}\\
B_{nl}^{-}=&\frac{(-1)^n}{(n+|l|)!}[\bar m |\beta|^{2}]^n 
\left\{
 \begin{array}{c}
  [- (\bar m+1)\beta^\ast]^{|l|}\\
  \left[\bar m \beta\right]^l
 \end{array}
\right\}\nonumber\\*
&e^{-i\bar m \Im \beta^2}e^{-\left(\bar m+\frac{1}{2}\right) |\beta|^2}\label{eq:Bminus}
\end{align}%
}%
from the generating function~\eqref{eq:wB} by differentiation with respect to $z$ and setting $z\to0$ afterwards.

\subsection{Overlap of displaced eigenelements}
\label{app:C}

The evaluation of Eq.~\eqref{eq:appCnlmk} goes along the lines sketched in App.~\ref{app:B}, this time, however, using the displaced right eigenelement, Eq.~\eqref{eq:muhatosc}, with the function
\begin{align}
 Q_{mk}(\alpha) = \frac1\pi\bra{\alpha}D(\beta)\hat\mu_{mk}(b,b^\dagger)D^\dagger(\beta)\ket{\alpha}.
\end{align}
The generating function, Eq.~\eqref{eq:lgf}, is applied twice to express the Laguerre polynomials in $\check{P}_{nl}$ and $Q_{mk}$ in terms of exponentials. The resulting integrals can be solved using
\begin{align}
 \int {\rm d}^2\alpha& \left\{\begin{array}{c}\alpha^{\nu}(\alpha^\ast+\beta^\ast)^\mu\\(\alpha+\beta)^\nu\alpha^{\ast\mu}\end{array}\right\} e^{\alpha\zeta^\ast-\alpha^\ast\xi}e^{-c|\alpha|^2}  \nonumber\\
 &=\frac{\pi}{c}e^{-\frac{\xi\zeta^\ast}{c}}
 \left\{
  \begin{array}{c}
   c^{-\nu} (-\xi)^{\nu-\mu} \mu! \,\text{L}_\mu^{\nu-\mu}[\xi(\beta^\ast+\zeta^\ast/c)]\\
   c^{-\mu} \zeta^{\ast\mu-\nu} \nu! \,\text{L}_\nu^{\mu-\nu}[-\zeta^\ast(\beta-\xi/c)]
  \end{array}
 \right\}.\label{eq:intform2}
\end{align}
From differentiating the generating function then follows
\begin{align}
C_{nl}^{mk}
 &=\binom{n}{m}\left[\frac{|\beta|^2}{\bar m+1}\right]^{n-m}\times\nonumber\\
 &\left\{
  \begin{array}{cl}\displaystyle
   \frac{1}{(n-m+|l|-|k|)!}\beta^{|l|-|k|}\quad&\text{for }l\ge0,\;k\ge0\\\displaystyle
  \frac{(n-m)!}{(n-m-|k|)!(n-m+|l|)!}\frac{\beta^{\ast|l|}}{\beta^{|k|}}\quad&\text{for }l<0,\;k\ge0\\
  \end{array}
 \right.\label{eq:Crel}
\end{align}
with the complex conjugate expressions when both $l$ and $k$ change sign.

\end{appendix}




\bibliography{article2}

\end{document}